\documentclass[british]{iopart}
\usepackage[T1]{fontenc}
\usepackage[latin9]{inputenc}
\usepackage{color}
\usepackage{babel}
\usepackage{float}
\usepackage{amsbsy}
\usepackage{amstext}
\usepackage{amssymb}
\usepackage{graphicx}
\usepackage{esint}
\usepackage[unicode=true,pdfusetitle,
 bookmarks=true,bookmarksnumbered=false,bookmarksopen=false,
 breaklinks=true,pdfborder={0 0 0},backref=false,colorlinks=true]
 {hyperref}
\usepackage{breakurl}

\makeatletter

\providecommand{\tabularnewline}{\\}

\usepackage{iopams}
\usepackage{setstack}

\newcommand{\eqref}[1]{(\ref{#1})}
\definecolor{darkgreen}{rgb}{0.423,0.643,0.698}
\definecolor{darkblue}{rgb}{0.773,0.325,0.476}
\definecolor{darkyellow}{rgb}{0.860,0.709,0.470}
\usepackage{doi}

\makeatother

\begin{document}

\title{Control of transversal instabilities in reaction-diffusion systems}

\author{Sonja Molnos}

\author{Jakob Löber}

\ead{jakob@physik.tu-berlin.de}

\author{Jan Frederik Totz}

\author{Harald Engel}

\address{Institut für Theoretische Physik, EW 7-1, Hardenbergstraße 36, Technische
Universität Berlin, 10623 Berlin, Germany}
\begin{abstract}
In two-dimensional reaction-diffusion systems, local curvature perturbations
in the shape of traveling waves are typically damped out and disappear
in the course of time. If, however, the inhibitor diffuses much faster
than the activator, transversal instabilities can arise, leading from
flat to folded, spatio-temporally modulated wave shapes and to spreading
spiral turbulence. For experimentally relevant parameter values, the
photosensitive Belousov-Zhabotinsky reaction (PBZR) does not exhibit
transversal wave instabilities. Here, we propose a mechanism to artificially
induce these instabilities via a wave shape dependent spatio-temporal
feedback loop, and study the emerging wave patterns. In numerical
simulations with the modified Oregonator model for the PBZR using
experimentally realistic parameter values we demonstrate the feasibility
of this control scheme. Conversely, in a piecewise-linear version
of the FitzHugh-Nagumo model transversal instabilities and spiral
turbulence in the uncontrolled system are shown to be suppressed in
the presence of control, thereby stabilising flat wave propagation.
\end{abstract}


\maketitle

\section{Introduction}

A large variety of pattern forming processes can be understood in
terms of the advancement of an interface between two or more spatial
domains. An interface that becomes unstable to diffusion possibly
causes intricate spatio-temporal dynamics. Well known examples include
the Mullins-Sekerka instability during crystal growth and formation
of snow flakes \cite{langer1980instabilities,brener1991pattern},
and the Saffman-Taylor instability leading to viscous fingering in
multiphase flow and porous media \cite{saffman1958penetration,saarloos1998three,de1999viscous}.
Other phenomena affected by interfacial instabilities are flame fronts
\cite{sivashinsky1977nonlinear,zeldovichmathematical}, Marangoni
convection \cite{birikh2003liquid}, and growing cell monolayers \cite{mehes2014collective}.\\
Traveling plane waves in excitable media exhibit interfacial instabilities
as well. Here, an effective interface separates the excited state
from the excitable rest state. A straight isoconcentration line of
a two-dimensional flat wave can suffer an instability leading to stationary
or time dependent modulations orthogonal to the propagation direction.
Further away from the instability threshold, rotating wave segments
and spreading spiral turbulence are observed \cite{maree1997spiral,zykov1998wave}.
For standard activator-inhibitor kinetics, these so-called transversal
or lateral wave instabilities typically occur if the inhibitor diffuses
much faster than the activator. This result was analytically predicted
first by Kuramoto for piecewise-linear reaction kinetics \cite{kuramoto2003chemical,kuramoto1980instability}.
Later, it was confirmed numerically by Horv{á}th et al. for autocatalytic
reaction-diffusion fronts with cubic reaction kinetics \cite{horv1993instabilities}
as well as in experiments with the iodate-arsenous acid reaction \cite{horvath1995instabilities}
and the acid-catalyzed chlorite\textendash{}tetrathionate reaction
\cite{horvath1998diffusion}.\\
The experimental workhorse of chemical pattern formation, the Belousov-Zhabotinsky
(BZ) reaction, does typically not display transversal wave instabilities.
Dispersing the reagents of the BZ reaction in nanodroplets of a water-in-oil
microemulsion allows to increase the inhibitor diffusivity considerably
\cite{vanag2001pattern} and leads, for example, to segmented spiral
waves as reported by Vanag and Epstein \cite{vanag2003segmented}.
Even in the presence of an electrical field applied to enhance transversal
instabilities in cubic autocatalytic reaction-diffusion fronts, the
inhibitor diffusion coefficient is always required to be sufficiently
larger than that of the activator \cite{horvath1999electric,toth1999lateral,toth2001lateral}.\\
Because of the possibility to apply spatio-temporal external forcing
or feedback-mediated control loops by exploiting the dependence of
the local excitation threshold on the intensity of applied illumination,
the photosensitive variant of the BZ reaction has been widely used
as a paradigm of an experimentally well controllable RD system. So
far, unstable wave propagation has been stabilised by global feedback
\cite{mihaliuk2002feedback}. Two feedback loops were used to stabilise
unstable wave segments and to guide their propagation along pre-given
trajectories \cite{sakurai2002design}. Also, spiral wave drift in
response to resonant external forcing and various feedback-mediated
control loops has been extensively studied experimentally in PBZR
systems, compare for example \cite{steinbock1993control,zykov1994external,PhysRevLett.92.018304,schlesner2006stabilization,schlesner2008efficient}.\\
In this paper, we design a curvature-dependent spatio-temporal feedback
loop in order to destabilise a stable propagating planar reaction-diffusion
wave by inducing transversal instabilities. In numerical simulations
with the modified Oregonator model for the PBZR, we study the wave
patterns emerging beyond the instability threshold, and demonstrate
the capability to actively select wave patterns by modifying feedback
parameters accessible to the experimenter. Conversely, under conditions
where planar wave propagation fails due to transversal instabilities,
using the same feedback mechanism we suppress ongoing breakup and
segmentation of waves, thereby stabilising unstable propagating planar
waves.

\section{Theory}

\subsection{Models}

The first model we investigate is the three component modified Oregonator
model \cite{krug1990analysis} 
\begin{eqnarray}
\frac{\partial u}{\partial t} & = & \frac{1}{\epsilon}\left[u-u^{2}+w\left(q-u\right)\right]+D_{u}\Delta u,\label{eq:OregonatorActivator}\\
\frac{\partial v}{\partial t} & = & u-v,\label{eq:OregonatorCatalyst}\\
\frac{\partial w}{\partial t} & = & \frac{1}{\tilde{\epsilon}}\left[\Phi+fv-w\left(u+q\right)\right]+D_{w}\Delta w.\label{eq:OregonatorInhibitor}
\end{eqnarray}
Here, the parameters $\epsilon$ and $\tilde{\epsilon}$ characterise
the time scales for the dynamics of the activator $u$ and inhibitor
$w$, respectively, and the stoichiometric parameters $q$ and $f$
depend on the temperature and chemical composition. All parameter
values used in numerical simulations are listed in table \ref{tab:3cOregonatorModelTab}
in \ref{sec:AppendixA}. The modified Oregonator model describes the
light-sensitive Belousov-Zhabotinsky (BZ) reaction. In experiments,
the catalyst $v$ can be immobilised in a gel and therefore the corresponding
diffusion coefficient is set to zero. The activator $u$ and inhibitor
$w$ diffuse with diffusion coefficients $D_{u}$ and $D_{w}$, whose
ratio for typical BZ recipes is approximately $D_{w}/D_{u}\approx1.2$.
This value is too low to support transversal instabilities such that
plane waves are stable for typical BZ recipes. The parameter $\Phi$
in equation \eqref{eq:OregonatorInhibitor} is proportional to the
applied light intensity and measures the local excitation threshold.
In experiments, spatio-temporal modulations of $\Phi$ can be applied
to control wave propagation in the BZ reaction \cite{sakurai2002design,schlesner2006stabilization,schlesner2008efficient,mihaliuk2002feedback}.\\
Because the modified Oregonator model does not exhibit transversal
instabilities in the parameter regime relevant for experiments, we
investigate a second model. The piecewise-linear caricature of the
FitzHugh-Nagumo (FHN) model \cite{zykov1987simulation,zykov1998wave}
received some attention in the context of transversal instabilities
\cite{zykov1998wave}. It is a two component model of standard activator-inhibitor
type,
\begin{figure}[t]
\begin{center}\includegraphics[width=1\textwidth]{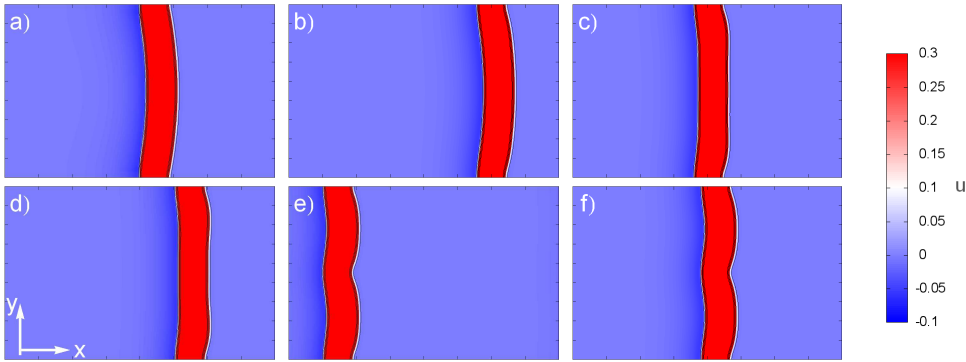}\end{center}
\caption{\label{fig:FHNCloseToOnset}Slightly beyond the onset of transversal
instabilities, an initially plane wave in the piecewise linear FitzHugh-Nagumo
model develops a stationary fold. From left to right, top to bottom:
time sequence of snapshots of a video which can be found in the Supplemental
Material \cite{supplement} for (a) $t=100$, (b) $t=320$, (c) $t=370$,
(d) $t=420$, (e) $t=470$, and (f) $t=570$ for the parameters $\sigma=2.1,\,\epsilon=0.1425,\, a=0.1$.
Computations were carried out with a constant time step $\Delta t=0.00001$
and space step $\Delta x=0.15$. The domain size is $160\times90$.}

\label{transversalinstabilitiesFHN1} 
\end{figure}
 
\begin{eqnarray}
\frac{\partial u}{\partial t} & = & \triangle u+F\left(u,v\right),\label{eq:PiecewiseLinearFHNActivator}\\
\frac{\partial v}{\partial t} & = & \sigma\triangle v+\epsilon G\left(u,v\right),\label{eq:PiecewiseLinearFHNInhibitor}
\end{eqnarray}
with $u$ being the activator and $v$ the inhibitor. The reaction
kinetics are a piecewise-linear caricature of the FHN model 
\begin{eqnarray}
F\left(u,v\right) & = & f\left(u\right)-v,\\
G\left(u,v\right) & = & k_{g}u-v,
\end{eqnarray}
where
\begin{equation}
f\left(u\right)=\left\{ \begin{array}{@{}l@{\quad}l}
-k_{1}u, & u<\delta,\\
k_{f}\left(u-a\right), & \delta<u<1-\delta,\\
k_{2}\left(1-u\right), & 1-\delta<u.
\end{array}\right.
\end{equation}
The parameters $k_{1}$ and $k_{2}$ are chosen such that $f\left(u\right)$
is continuous at $u=\delta$ and $u=1-\delta$, which leads to
\begin{eqnarray}
k_{1} & = & \frac{k_{f}}{\delta}\left(a-\delta\right),\qquad k_{2}=\frac{k_{f}}{\delta}\left(1-\delta-a\right).
\end{eqnarray}
The remaining parameters for the function $f$ are chosen in such
a way that $f$ resembles the cubic shape of the FHN activator nullcline.
All parameter values used in numerical simulations are listed in table
\ref{tab:RinzelKellertab} in \ref{sec:AppendixA}. The parameter
$a$ is a measure for the excitation threshold and used as the feedback
parameter.\\
For numerical simulations, we assume an elongated two-dimensional
channel of width $L$ in the $y$-direction with waves propagating
in the $x$-direction. The boundary conditions in the $x$-direction
are periodic while we assume periodic or Neumann boundary conditions
in the $y$-direction. For both models, we use a box-like initial
condition of width $b$ for the vector of components $\mathbf{u}$,

\begin{eqnarray}
\mathbf{u}\left(x,y,t_{0}\right) & = & \Theta_{\text{Box}}\left(\left(x-\phi\left(y,t_{0}\right)\right)/b\right)\left(\mathbf{u}_{\text{max}}-\mathbf{u}_{0}\right)+\mathbf{u}_{0},\label{eq:InitialCondition}
\end{eqnarray}
where $\mathbf{u}_{\text{max}}$ is the initial height of the pulse
and $\mathbf{u}_{0}$ is the stationary point of the reaction kinetics.
The box function is defined as
\begin{equation}
\Theta_{\text{Box}}\left(x\right)=\left\{ \begin{array}{@{}l@{\quad}l}
1, & \left|x\right|<1/2,\\
0, & \left|x\right|\geq1/2.
\end{array}\right.
\end{equation}
The initial shape of the wave is given by
\begin{eqnarray}
\phi\left(y,t_{0}\right) & = & d-A\cos\left(\frac{2\pi y}{L}\right)\text{,}\label{eq:InitialShape}
\end{eqnarray}
where $A$ denotes the amplitude of deviation of the shape from a
plane wave and $d$ is an offset. For numerical simulations in two
spatial dimensions, we use Euler forward for the time evolution and
a five point stencil for the Laplacian.\\
A phase diagram for the occurrence of transversal instabilities in
the $\epsilon$-$\sigma$-parameter plane of the piecewise-linear
FHN model was presented by Zykov et al. in \cite{zykov1998wave}.
Increasing the inhibitor diffusion coefficient $\sigma$ crosses the
threshold for transversal instabilities. Shortly beyond the onset
of transversal instabilities, a plane wave develops a fold which is
stationary in a comoving frame of reference, see figure \ref{fig:FHNCloseToOnset}
for a time sequence of snapshots and the supplemental material \cite{supplement}
for a movie. Further away from the instability threshold, a plane
wave breaks into segments which undergo self-sustained rotatory motion
accompanied by permanent merging and annihilation of segments. This
regime is also known as spreading spiral turbulence \cite{zykov1998wave},
see figure \ref{fig:FHNFarFromOnset} for a time sequence of snapshots.
\begin{figure}[b]
\begin{center}\includegraphics[width=1\textwidth]{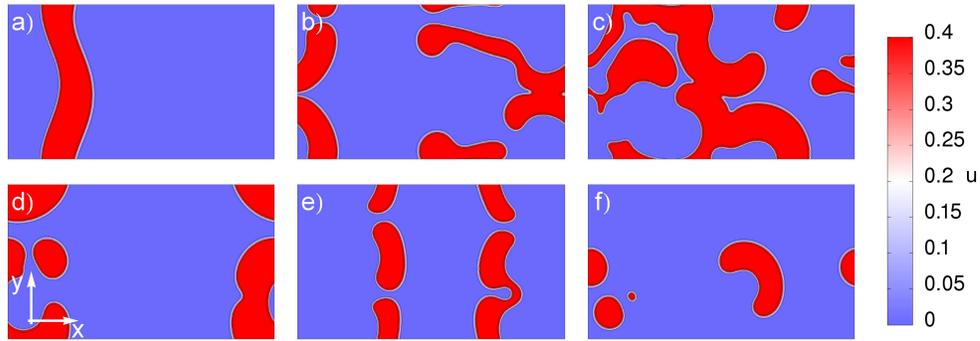}\end{center}
\caption{\label{fig:FHNFarFromOnset}Segmentation of initially plane waves
and spreading spiral turbulence occurs deep in the regime of transversal
instabilities in the piecewise linear FitzHugh-Nagumo model. Parameter
values as in figure \ref{fig:FHNCloseToOnset} except $\epsilon=0.1575$.
The segments undergo self-sustained rotatory motion and nucleate new
waves. From left to right, top to bottom: snapshots of a video \cite{supplement}
for (a) $t=5$, (b) $t=301$, (c) $t=500$, (d) $t=672$, (e) $t=700$,
and (f) $t=891$. The domain size is $160\times90$.}
\end{figure}
\\

\subsection{Evolution equations for isoconcentration lines\label{sub:EvolutionEquations}}

Theoretically, the onset of transversal instabilities can be understood
with the help of the linear eikonal equation
\begin{eqnarray}
c_{n}\left(s,t\right) & = & c-\nu\kappa\left(s,t\right),\label{eq:LinearEikonalEquation}
\end{eqnarray}
an evolution equation for a two-dimensional curve $\boldsymbol{\gamma}\left(s,t\right)=\left(\gamma_{x}\left(s,t\right),\gamma_{y}\left(s,t\right)\right)^{T}$
representing an isoconcentration line parametrised by the curve parameter
$s$. The linear eikonal equation relates the normal velocity ($\boldsymbol{n}$
is the normal vector of $\boldsymbol{\gamma}$) 
\begin{eqnarray}
c_{n}\left(s,t\right) & = & \boldsymbol{n}\left(s,t\right)\cdot\partial_{t}\boldsymbol{\gamma}\left(s,t\right)
\end{eqnarray}
along $\boldsymbol{\gamma}$ linearly to its curvature,
\begin{eqnarray}
\kappa\left(s,t\right) & = & \frac{\partial_{s}\gamma_{x}\left(s,t\right)\partial_{s}^{2}\gamma_{y}\left(s,t\right)-\partial_{s}\gamma_{y}\left(s,t\right)\partial_{s}^{2}\gamma_{x}\left(s,t\right)}{\left(\left(\partial_{s}\gamma_{x}\left(s,t\right)\right)^{2}+\left(\partial_{s}\gamma_{y}\left(s,t\right)\right)^{2}\right)^{3/2}}.\label{eq:Curvature}
\end{eqnarray}
The curvature is conventionally assumed to be positive for convex
isoconcentration lines, i.e., an isoconcentration lines with a protrusion
in the propagation direction. The constant $c$ corresponds to the
pulse velocity of a one-dimensional solitary wave and $\nu$ is the
curvature coefficient. A rigorous derivation of the eikonal equation
\eqref{eq:LinearEikonalEquation} from the reaction-diffusion system
identifies the constant $\nu$ in terms of the one-dimensional pulse
profile, its response function and the matrix of diffusion coefficients,
see \cite{dierckx2011accurate} for details. For a plane wave, any
isoconcentration level is a straight line and therefore its curvature
vanishes, $\kappa\left(s,t\right)\equiv0$, everywhere along $\boldsymbol{\gamma}$.
The stability of a plane wave is determined by the sign of the curvature
coefficient $\nu$. As long as $\nu>0$, any point along the isoconcentration
line of a convex bulge moves slower than a plane wave, while points
of a concave dent move faster than a plane wave, thereby smoothing
out deviations from a plane wave. If $\nu<0$, a convex bulge moves
faster than a plane wave, protruding the bulge even further and thereby
leading to an ever increasing curvature: a transversal instability
arises. Patterns arising for $\nu<0$ cannot be described by the linear
eikonal equation and terms depending nonlinearly on the curvature
have to be taken into account which saturate the growth of an ever
increasing curvature. At least two different nonlinear versions of
equation \eqref{eq:LinearEikonalEquation} exist in the literature.
Zykov et al. \cite{zykov1980biofizika,zykov1987simulation,mikhailov1991kinematical,zykov2004feedback}
renormalised $\epsilon$ and $\sigma$ in equation \eqref{eq:PiecewiseLinearFHNInhibitor}
to derive a renormalised one-dimensional velocity $c$ depending on
the curvature. Dierckx et al. \cite{dierckx2011accurate} derive higher
order nonlinear corrections in the curvature by a rigorous perturbation
expansion with a small parameter proportional to the curvature, additionally
generalising the eikonal equation to anisotropic media.\\
Apart from nonlinear eikonal equations, which are difficult to solve
numerically, patterns arising beyond the threshold of transversal
instability can be described by the Kuramoto-Sivashinsky (KS) equation,

\begin{eqnarray}
\partial_{t}\phi\left(y,t\right) & = & c+\frac{c}{2}\left(\partial_{y}\phi\left(y,t\right)\right)^{2}+\nu\partial_{y}^{2}\phi\left(y,t\right)-\lambda\partial_{y}^{4}\phi\left(y,t\right).\label{eq:ksgl}
\end{eqnarray}
Equation \eqref{eq:ksgl} is an evolution equation for the $x$-component
$\phi\left(y,t\right)$ of an isoconcentration line $\boldsymbol{\gamma}$
parametrised in the form $\boldsymbol{\gamma}\left(y,t\right)=\left(\phi\left(y,t\right),y\right)^{T}$.
See \cite{kuramoto1980instability} and \cite{malevanets1995biscale}
for a derivation of equation \eqref{eq:ksgl} from a general RD system.
The case of Neumann boundary conditions in the $y$-direction for
the RD system imply that any isoconcentration line of activator and
inhibitor meets the domain boundary in a right angle. This corresponds
to Neumann boundary conditions for $\phi$, 
\begin{eqnarray}
\partial_{y}\phi\left(0,t\right) & =0,\, & \partial_{y}\phi\left(L,t\right)=0.
\end{eqnarray}
Similarly, periodic boundary conditions in the $y$-direction of the
RD system carry over to periodic boundary conditions for $\phi$.
Equation \eqref{eq:ksgl} was originally proposed by Sivashinsky \cite{sivashinsky1977nonlinear}
in the study of turbulent flame propagation and adapted for reaction-diffusion
systems by Kuramoto \cite{kuramoto1978diffusion,kuramoto1980instability}.
The parameter $\lambda$ can be expressed in terms of a sum over all
eigenfunctions of the linear stability operator arising through a
linearisation of the one-dimensional RD system around the traveling
wave solution \cite{kuramoto1980instability}. To compute $\lambda$,
we use a method which avoids the virtually impossible numerical computation
of all eigenfunctions, see \cite{malevanets1995biscale} for details.
The values of $\lambda$ and $\nu$ for the modified Oregonator model
with parameters as given in \ref{sec:AppendixA} are
\begin{eqnarray}
\lambda= & 0.68,\qquad & \nu=1.05.
\end{eqnarray}
The Kuramoto-Sivashinsky equation \eqref{eq:ksgl} allows a refined
investigation of the onset of transversal instabilities. For a stability
analysis of a plane wave in channel of width $L$ with Neumann boundary
conditions, we apply a perturbation expansion in $0<\hat{\epsilon}\ll1$
with an ansatz in form of a Fourier series, 
\begin{eqnarray}
\phi\left(y,t\right) & = & ct+\hat{\epsilon}\sum_{n=-\infty}^{\infty}a_{n}\exp\left(\omega_{n}t\right)\cos\left(\frac{\pi ny}{L}\right),\label{eq:ansatz_KSGL}
\end{eqnarray}
where the term $ct$ corresponds to a plane wave solution to the RD
system traveling in the $x$-direction. The dispersion relation follows
as 
\begin{eqnarray}
\omega_{n} & = & -\lambda\left(\frac{n\pi}{L}\right)^{4}-\nu\left(\frac{n\pi}{L}\right)^{2}.
\end{eqnarray}
Transversal instabilities occur only if $\omega_{1}>0$, i.e., $\nu$
must be negative and the channel width must exceed 
\begin{eqnarray}
L & = & \pi\sqrt{\frac{\lambda}{-\nu}}.\label{eq:ConditionOfTransversalInstability}
\end{eqnarray}
Thus, in general, the onset of a transversal instability depends on
the boundary conditions and can be suppressed in thin channels. It
is a long-wavelength instability, i.e., the first mode which becomes
unstable upon reaching the threshold is the mode with the longest
possible wavelength.\\
If $\nu<0$, the fourth order term in the KS equation \eqref{eq:ksgl}
counteracts the negative diffusion term and leads to a saturation
of the growth of wavefront modulations.Starting at the threshold of
instability, the solution to the KS equation\eqref{eq:ksgl} displays
a fold with a minimum located at $y=L/2$. Upon increasing $L$, this
steady wave loses stability via a supercritical Hopf bifurcation \cite{horv1993instabilities}
and the wave front starts to oscillate back and forth in a symmetrical
fashion. Increasing $L$ even further leads to a symmetry breaking
bifurcation with asymmetrical oscillations followed by a period doubling
cascade to fully developed spatio-temporal chaos. In this regime,
the KS equation displays a strong dependence on the initial data,
with small differences in the initial conditions leading to dramatically
different future time evolution. This characteristic of the KS equation
is also studied as an analogy for hydrodynamic turbulence \cite{ChaosBook}.\\
As long as $\nu>0$, no instability can arise and the fourth order
term can be safely neglected by setting $\lambda=0$. In this case,
equation \eqref{eq:ksgl} simplifies to the nonlinear phase diffusion
equation, which in turn can be transformed to the usual diffusion
equation via the Cole-Hopf transform \cite{kuramoto2003chemical}.
Therefore, equation \eqref{eq:ksgl} with $\lambda=0$ can be solved
analytically for arbitrary initial and boundary conditions.\\
To assess the accuracy of the Kuramoto-Sivashinsky equation \eqref{eq:ksgl}
as an approximation for propagating reaction-diffusion waves, we compare
the transition from an initially curved shape to a plane wave for
$\nu>0$ with numerical simulations of the underlying two-dimensional
modified Oregonator model Eqs. \eqref{eq:OregonatorActivator}-\eqref{eq:OregonatorInhibitor}.
The isoconcentration line $\boldsymbol{\gamma}$ of the activator
variable $u$ is determined numerically as the set of points $\mathbf{r}=\left(x,\, y\right)^{T}$
for which $u\left(\mathbf{r},t\right)=u_{c}=0.2$. We compute the
$x$-component of the wave's centre of mass in\textcolor{black}{{} a
comoving frame as}
\begin{figure}[t]
\begin{center}\includegraphics[scale=0.4]{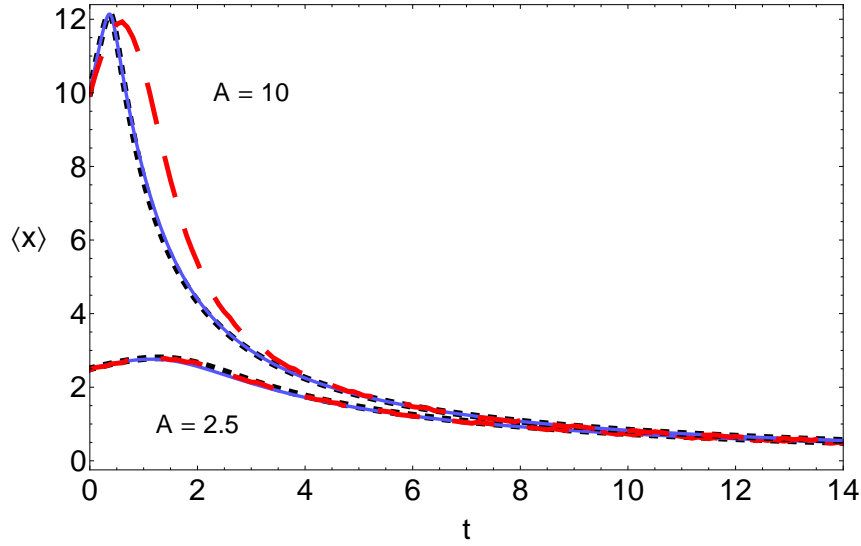}\end{center} \caption{\label{fig:TimeEvolutionMeanXValue}Time evolution of the mean $x$-coordinate
of an isoconcentration line\textcolor{black}{{} in a comoving frame
of reference}. Comparison of Kuramoto-Sivashinsky equation (black
dotted line), nonlinear phase-diffusion equation (blue line) and for
the activator isoconcentration line with $u\left(x,y,t\right)=0.2$
of the modified Oregonator obtained by numerical simulations (red
dotted line).}
\end{figure}
\begin{eqnarray}
\left\langle x\right\rangle \left(t\right) & = & \frac{1}{L}\intop_{0}^{L}\phi\left(y,t\right)dy-\phi\left(0,t\right).
\end{eqnarray}
Figure \ref{fig:TimeEvolutionMeanXValue} shows the time evolution
of $\left\langle x\right\rangle \left(t\right)$ obtained from the
Kuramoto-Sivashinsky equation (black dotted line) and nonlinear phase
diffusion equation (blue solid line) and for the modified Oregonator
model obtained by numerical simulations (red dashed line) for two
different values of the amplitude $A$ which characterises the initial
deviation from a plane wave. As one would expect intuitively, the
agreement between numerical simulations on the one hand and Kuramoto-Sivashinsky
equation and nonlinear phase diffusion equation on the other hand
becomes worse the larger is the initial amplitude $A$. For large
times, i.e., when the curved isoconcentration line approaches a straight
line, all results agree. The nonlinear phase diffusion equation and
the Kuramoto-Sivashinsky equation practically yield the same result
for all times, confirming the fact that the fourth order derivative
in the Kuramoto-Sivashinsky equation can safely be neglected if the
curvature coefficient is $\nu>0$.
\begin{figure}[t]
\begin{center}\includegraphics[scale=0.35]{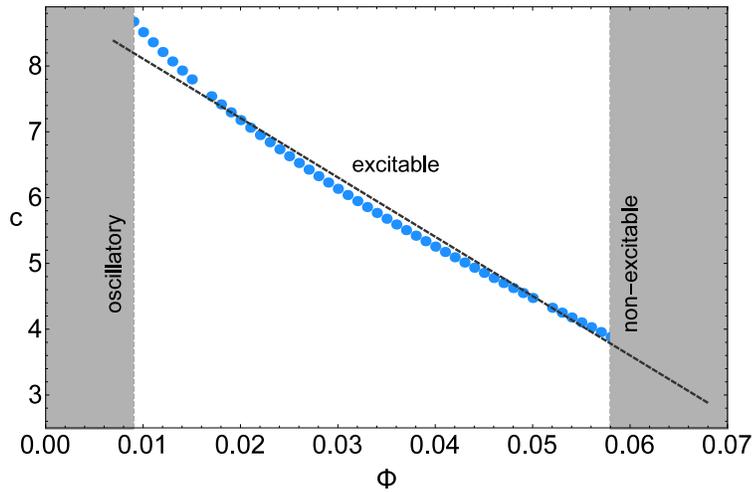}\end{center} \caption{\label{fig:c0OverPhi}Velocity $c$ of a one-dimensional solitary
pulse over the parameter $\Phi$ proportional to applied light intensity
for the modified Oregonator model. The result of numerical simulations
(blue dots) can be well approximated by a linear least square fit
(blue solid line). }
\end{figure}

\subsection{Curvature-dependent feedback control}

The feedback law proposed in this article requires that the velocity
$c$ of a one-dimensional wave depends sufficiently strongly on a
parameter which can be controlled in experiments. For the modified
Oregonator model, we use the parameter $\Phi$ proportional to the
applied light intensity as the feedback parameter. A numerical computation
of the dependence of $c$ on $\Phi$ is shown in figure \ref{fig:c0OverPhi}.
With relatively good accuracy, the dependence can be assumed to be
linear,

\begin{eqnarray}
c\left(\Phi\right) & = & c_{0}+c_{1}\Phi,\label{eq:c0OverPhi}
\end{eqnarray}
with parameters $c_{1}=-90.191,\, c_{0}=9.013$ obtained from a least
square fit. Solitary waves exist only in the excitable regime of $\Phi$
values indicated by the dashed lines in figure \ref{fig:c0OverPhi}.
For $\Phi\lesssim0.045$,  the rest state is unstable and the medium
becomes oscillatory. For $\Phi\gtrsim0.068$, the solitary pulse profile
becomes unstable and decays to the stable rest state. A successful
feedback control is possible if $\Phi$ is restricted to lie between
these two values.\\
\begin{figure}[t]
\centering \includegraphics[width=1\textwidth]{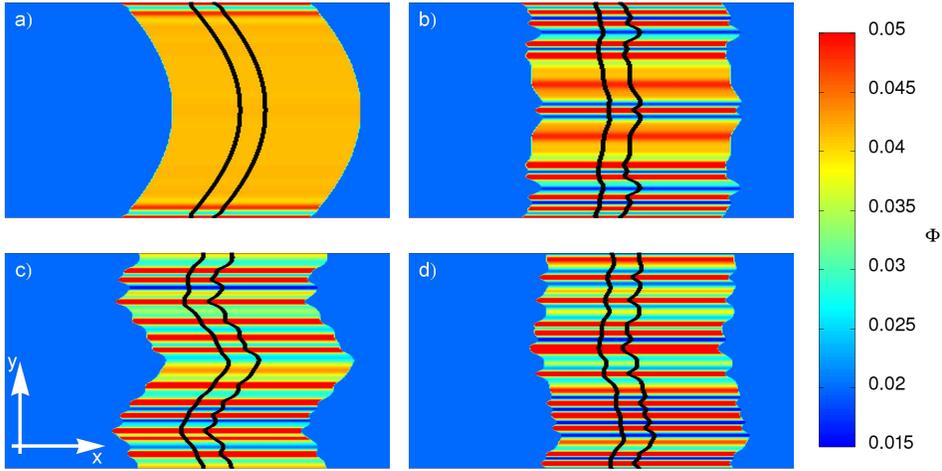} \caption{\label{fig:WeakFeedBack}Small modulations of the wave shape occur
for weak feedback with an effective curvature coefficient of $\tilde{\nu}=-0.75$.
Snapshots of a movie \cite{supplement} for (a) $t=3.5$, (b) $t=10.6$,
(c) $t=25.2$, and (d) $t=69.3$. Colours denote the value of the
applied spatio-temporal illumination field $\Phi\left(x,y,t\right)$
which is proportional to the curvature of the black isoconcentration
line. Shown are clippings of size $21.5\times30$ centred on the wave's
centre of mass while the computational domain is $110\times30$. This
and the following computations of the modified Oregonator model were
carried out with a time step size of $\Delta t=0.0001$ and a space
step size $\Delta x=0.05$. Parameter values for the feedback law
are $\Phi_{\text{min}}=0.018,\,\Phi_{\text{max}}=0.042$.}
\end{figure}
We introduce a feedback law for $\Phi$ depending linearly on the
curvature, 
\begin{eqnarray}
\Phi\left(\kappa\right) & = & \alpha+\beta\kappa.\label{eq:FeedbackLaw}
\end{eqnarray}
The parameters $\alpha$ and $\beta$ are accessible to an experimenter.
In general, these parameters can be adjusted with time to achieve
a better performance of the control. Together with equation \eqref{eq:c0OverPhi}
and equation \eqref{eq:FeedbackLaw}, the linear eikonal equation
\eqref{eq:LinearEikonalEquation} becomes 
\begin{eqnarray}
c_{n} & = & c_{1}\alpha+c_{0}-\tilde{\nu}\kappa,\label{eq:LinearEikonalEquationWithFeedback}
\end{eqnarray}
with the effective curvature coefficient
\begin{equation}
\tilde{\nu}=\nu-c_{1}\beta.
\end{equation}
Depending on the sign of $\tilde{\nu}$, the control will have very
different effects. If a plane wave is stable with respect to transversal
perturbations because $\nu>0$, we can excite transversal instabilities
if $\tilde{\nu}=\nu-c_{1}\beta<0$. Conversely, if $\nu>0$ such that
plane waves are unstable with respect to transversal modulations,
patterns can be stabilised if $\tilde{\nu}=\nu-c_{1}\beta>0$. An
appropriate choice of the parameters $\alpha$ and $\beta$ in the
feedback law \eqref{eq:FeedbackLaw} allows to control transversal
instabilities.\\
To apply the feedback law \eqref{eq:FeedbackLaw} it is necessary
to compute the curvature of a chosen isoconcentration line of a chosen
component with sufficient accuracy, which raises considerable difficulties.

\subsection{Computation of curvature by Level Set Methods\label{sub:ComputationOfCurvature}}

The curvature $\kappa\left(s,t\right)$ of an isoconcentration line
$\boldsymbol{\gamma}\left(s,t\right)$, equation \eqref{eq:Curvature},
is proportional to the second derivative of the isoconcentration line
with respect to the curve parameter $s$. Computations of isoconcentration
lines from numerical simulations or experiments are affected by noise
due to the discretised nature of the computed or measured concentration
field $u$, respectively. Numerical differentiation is an ill-posed
mathematical operation and typically amplifies the noise. A variety
of methods to compute the curvature $\kappa$ directly from a numerically
determined isoconcentration line were tested and discarded due to
insufficient performance.\\
An indirect method which avoids the differentiation of an isoconcentration
line is to compute the curvature field $\tilde{\kappa}$ as 
\begin{equation}
\tilde{\kappa}\left(\mathbf{r},t\right)=\nabla\mathbf{\cdot}\frac{\nabla u\left(\mathbf{r},t\right)}{\left|\nabla u\left(\mathbf{r},t\right)\right|}.\label{eq:curvaturefield}
\end{equation}
According to the formula of Bonnet \cite{abbena2006modern}, evaluating
$\tilde{\kappa}$ at an isoconcentration line $\mathbf{r}=\boldsymbol{\gamma}\left(s,t\right)$
of $u$ yields the curvature $\kappa$ of $\boldsymbol{\gamma}$,
i.e.,
\begin{eqnarray}
\kappa\left(s,t\right) & = & \tilde{\kappa}\left(\boldsymbol{\gamma}\left(s,t\right),t\right).
\end{eqnarray}
See \ref{sec:AppendixB} for a proof of Bonnet's formula. Equation
\eqref{eq:curvaturefield} involves the determination of the second
derivative of $u$ with respect to $x$ and $y$. These expressions
are readily available from the finite difference algorithm used to
solve the RD system numerically. The problem is now that the concentration
$u$ of a pulse solution typically varies very fast in a small spatial
region while it is constant everywhere else, leading to an ill-defined
denominator in equation \eqref{eq:curvaturefield}. This difficulty
can be addressed with the help of a level set method, which, however,
is numerically quite expensive.\\
\begin{figure}[t]
\centering \includegraphics[width=1\textwidth]{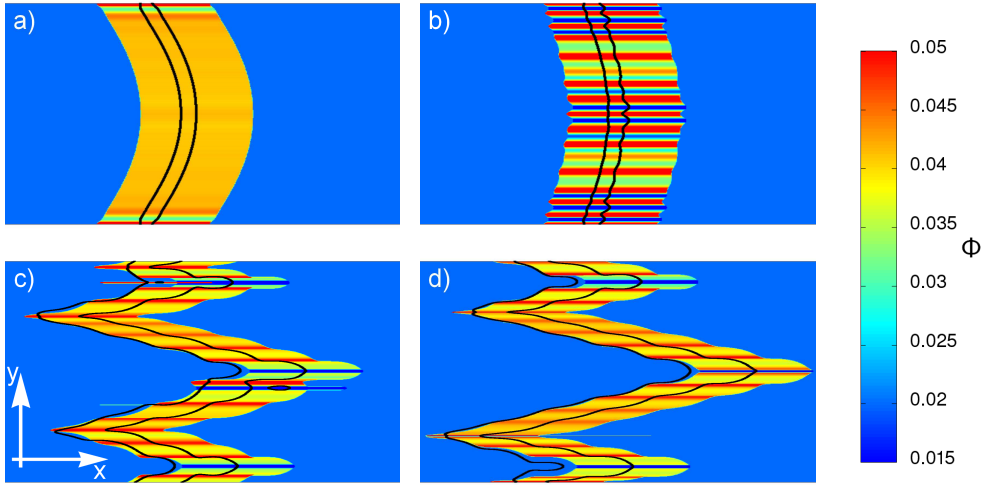} \caption{\label{fig:ModerateFeedBack}Moderate feedback corresponding to an
effective curvature coefficient of $\tilde{\nu}=-1.059$ leads to
V-shaped patterns moving much faster than a plane wave. Snapshots
of a movie \cite{supplement} for (a) $t=2$, (b) $t=7$, (c) $t=62$,
and (d) $t=80$. Shown are clippings of size $30\times30$ from a
computational domain of size $110\times30$, and feedback parameter
values are $\Phi_{\text{min}}=0.018,\,\Phi_{\text{max}}=0.046$.}
\end{figure}
Originally, level set methods were developed by Osher and Sethian
to compute and track the motion of interfaces. These methods have
since been successfully applied in such diverse areas of applications
as computer graphics, medical image segmentation and crystal growth
\cite{Osher198812,sethian1999level,osher2003level}.\\
We introduce a second field variable $\chi\left(\mathbf{r},\tau\right)$
which evolves in (virtual) time $\tau$ according to the so-called
reinitialisation equation \cite{russo2000remark,osher2003level,du2008second}
\begin{eqnarray}
\partial_{\tau}\chi+\text{sign}\left(\chi^{0}\right)\left(\left|\nabla\chi\right|-1\right) & = & 0\label{eq:ReinitializationEquation}
\end{eqnarray}
with 
\begin{equation}
\text{sign}\left(x\right)=\left\{ \begin{array}{@{}l@{\quad}l}
\hspace{3mm}1 & \mbox{if \ensuremath{x>0}}\\
\hspace{3mm}0 & \mbox{if \ensuremath{x=0}}\\
-1 & \mbox{if \ensuremath{x<0}}
\end{array}\right..
\end{equation}
Equation \eqref{eq:ReinitializationEquation} is solved with the initial
condition
\begin{eqnarray}
\chi\left(\mathbf{r},0\right) & =\chi^{0}\left(\mathbf{r}\right)= & u\left(\mathbf{r},t\right)-u_{c},
\end{eqnarray}
where $u_{c}$ is the activator value along the isoconcentration line
$\boldsymbol{\gamma}$ for which we want to determine the curvature
$\kappa$, i.e., $u\left(\boldsymbol{\gamma}\left(s,t\right),t\right)=u_{c}$.
Note that $\chi\left(\boldsymbol{\gamma}\left(s,t\right),\tau\right)=\chi^{0}\left(\boldsymbol{\gamma}\left(s,t\right)\right)=0$
for all times $\tau$ such that the position of the level set $\boldsymbol{\gamma}$
is not changed by equation \eqref{eq:ReinitializationEquation}. However,
equation \eqref{eq:ReinitializationEquation} transforms the neighbourhood
of $\chi=0$ such that, after sufficiently many time steps $\tau$,

\begin{equation}
\lim_{\tau\rightarrow\infty}\left|\nabla\chi\left(\mathbf{r},\tau\right)\right|=1.
\end{equation}
The curvature $\kappa$ of $\boldsymbol{\gamma}$, equation\eqref{eq:Curvature}
can now readily be computed in terms of the Laplacian of $\chi$ as

\begin{equation}
\kappa\left(s,t\right)=\tilde{\kappa}\left(\boldsymbol{\gamma}\left(s,t\right),t\right)=\lim_{\tau\rightarrow\infty}\Delta\chi\left(\boldsymbol{\gamma}\left(s,t\right),\tau\right).
\end{equation}
Numerically, the evolution of $\chi$ up to the final time $\tau=0.01$
is sufficient to obtain a very accurate smooth result for the curvature
of $\boldsymbol{\gamma}$. The reinitialisation equation \eqref{eq:ReinitializationEquation}
has to be solved at every real time step $t$. However, because the
time evolution of the RD system is slow enough, we recompute the curvature
$\kappa$ only every 200th time step $t$.
\begin{figure}[t]
\centering \includegraphics[width=1\textwidth]{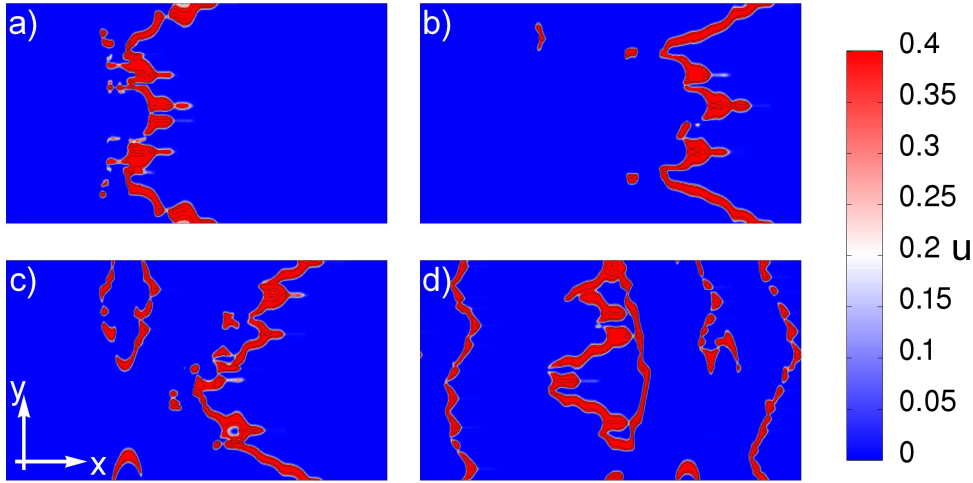} \caption{\label{fig:StrongFeedBack}Segmentation of waves occurs under very
strong feedback with an effective curvature coefficient of $\tilde{\nu}=-2.337$.
Segments may break off, nucleate new waves and often start to rotate.
Snapshots of a movie \cite{supplement} for (a) $t=4$, (b) $t=6$,
(c) $t=25$, and (d) $t=50$. The domain size is $110\times30$, and
feedback parameters are $\Phi_{\text{min}}=0.006,\,\Phi_{\text{max}}=0.05$.}
\end{figure}

\section{Results}

\subsection{Excitation of transversal instabilities in the modified Oregonator
model}

We study the possibility to excite transversal instabilities by curvature-dependent
feedback in the modified Oregonator model. The feedback law \eqref{eq:FeedbackLaw}
is realised via the parameter $\Phi$ proportional to the illumination
in the BZ reaction. For the parameters of the feedback law we set
$\alpha=\Phi_{\text{max}}$ and $\beta=-\left(\Phi_{\text{max}}-\Phi_{\text{min}}\right)/\kappa_{\text{norm}}$
such that the effective curvature coefficient is 
\begin{eqnarray}
\tilde{\nu} & = & \nu-c_{1}\beta=\nu+\frac{c_{1}}{\kappa_{\text{norm}}}\left(\Phi_{\text{max}}-\Phi_{\text{min}}\right).\label{eq:EffectiveCurvatureCoefficient}
\end{eqnarray}
The values of $\Phi_{\text{max}}$ and $\Phi_{\text{min}}$ can be
chosen arbitrarily as long as $\Phi_{\text{min}}<\Phi_{\text{max}}$
and both values lie in the regime of an excitable medium, see figure
\eqref{fig:c0OverPhi}. The curvature $\kappa$ is determined for
the activator isoconcentration line $\boldsymbol{\gamma}$ with $u\left(\boldsymbol{\gamma}\left(s,t\right),t\right)=u_{c}=0.2$.
An area of fixed size in front of and behind $\boldsymbol{\gamma}$
is illuminated with the same value $\Phi\left(\kappa\left(s,t\right)\right)$,
while within the remaining medium $\Phi$ attains its background value
$\Phi=\Phi_{0}$. Before the feedback is switched on at time $t_{1}=0.4$,
the wave evolves uncontrolled. The value of $\kappa_{\text{norm}}=1.2$
is an estimate of the largest value which the curvature attains during
the overall time evolution. For simplicity, we choose a constant value
of $\kappa_{\text{norm}}$, but in principle this value can be set
to the maximum curvature every time the curvature is recomputed.\\
Figure \ref{fig:WeakFeedBack} shows wave patterns arising for weak
feedback with an effective curvature coefficient $\tilde{\nu}=-0.75$.
The black solid lines denote the isoconcentration line $\boldsymbol{\gamma}$
for the activator level $u_{c}=0.2$. The rightmost line corresponds
to the wave front while the trailing line corresponds to the wave
back. The colours represent the value of the feedback parameter $\Phi$
and are proportional to the curvature of the wave front isoconcentration
line. An initially sinusoidal shape decays and a plane wave with transversal
modulations of small wavelength develops. For the example presented
here, the modulations are not stationary but travel along the isoconcentration
line until they annihilate each other or at the Neumann boundaries.
For even weaker feedback, the modulations do not travel such that
the pattern is truly stationary in a comoving frame of reference.
The overall velocity of the patterns is approximately the velocity
$c$ of the one-dimensional unperturbed traveling wave. Apart from
the wave length of the modulations, this type of pattern appears similar
to the patterns arising in the uncontrolled FHN model slightly beyond
the threshold of transversal instabilities, see figure \ref{fig:FHNCloseToOnset}.
\begin{figure}[t]
\centering \includegraphics[width=0.75\textwidth]{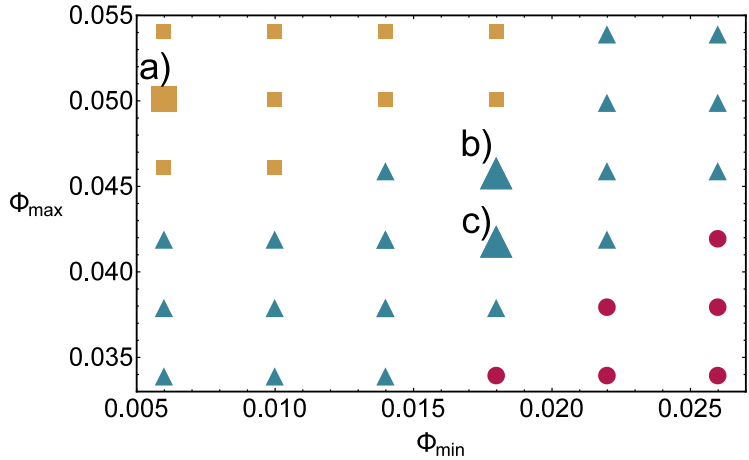} \caption{\label{fig:PhaseDiagram}Phase diagram for patterns in the modified
Oregonator model under feedback control. Blue bullets ${\color{darkblue}{\bullet}}$
correspond to folded waves stationary in a comoving frame of reference,
green triangles ${\color{darkgreen}{\blacktriangle}}$ stand for non-stationary
wave front modulations, ochre squares ${\color{darkyellow}{\blacksquare}}$
denotes segmentation of waves and spiral turbulence. a) refers to
figure \eqref{fig:StrongFeedBack}, b) to figure \eqref{fig:ModerateFeedBack}
and c) to figure \eqref{fig:WeakFeedBack}.}
\end{figure}
\\
Figure \ref{fig:ModerateFeedBack} displays the effects of moderate
feedback with an effective curvature coefficient $\tilde{\nu}=-1.059$.
V-shaped patterns arise which travel much faster than a corresponding
one-dimensional solitary pulse. In a frame of reference comoving with
the centre of mass, the V-shaped patterns appear stationary apart
from modulations traveling along the isoconcentration line. The V-patterns
observed under feedback are long-time stable and do not decay or break
up. A solitary V-pattern in an unbounded domain can be explained analytically
as a solution to the linear and nonlinear eikonal equations \cite{Brazhnik1995nonspiral,Brazhnik1996exact}.
A V with opening angle $\alpha$ has a mean velocity $V$ given by
\begin{figure}
\centering \includegraphics[width=0.6\textwidth]{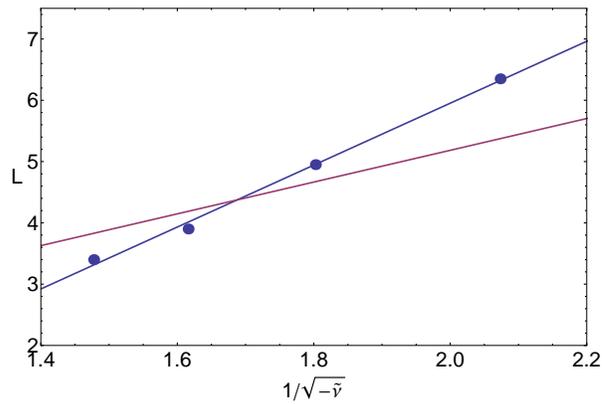} \caption{\label{fig:StabilityThreshold}Transversal instabilities are suppressed
in thin channels. Shown is the stability boundary of a plane wave
under feedback in a channel of width $L$ with Neumann boundary conditions.
Using feedback, the effective curvature coefficient $\tilde{\nu}=\nu-c_{1}\beta$
is adjusted until a plane wave becomes unstable. Red line: theoretical
prediction equation \eqref{eq:ConditionOfTransversalInstability}
obtained from the Kuramoto-Sivashinsky equation. Blue dots: result
of numerical simulations of the modified Oregonator model. Blue line:
least square fit to numerical results.}
\end{figure}
 
\begin{eqnarray}
V & = & \frac{c}{\sin\left(\alpha\right)},
\end{eqnarray}
where $c$ is the one-dimensional velocity. Because $\left|\sin\left(\alpha\right)\right|<1$,
all V-patterns are moving faster than a plane wave. Experimentally,
these patterns were observed in homogeneous \cite{Munuzuri1995Vshape}
and stratified \cite{Steinbock1993wave} BZ media.\\
Figure \ref{fig:StrongFeedBack} shows the effect of strong feedback
with an effective curvature coefficient $\tilde{\nu}=-2.337$. Similar
as for moderate feedback, V-shaped patterns appear. However, their
shape is non-stationary but oscillating. The V-shape is segmented
in an irregular and non-stationary way, with segments either merging
again or breaking off and serving as the nucleation centre for new
waves. These new waves propagate as segmented circles and occasionally
start to rotate until they annihilate upon collision with other waves.
Qualitatively, the segmentation and occurrence of rotating segments
is similar to the spreading spiral turbulence observed for the uncontrolled
FHN model deep in the regime of transversal instabilities, see figure
\ref{fig:FHNFarFromOnset}.\\
These results show that the proposed feedback law is not only able
to excite transversal instabilities but allows, to some extent, the
selection of the patterns beyond the instability threshold by tuning
the feedback parameters $\Phi_{\text{max}}$ and $\Phi_{\text{min}}$
accessible to an experimenter. We display a phase diagram with a classification
of the observed patterns in the $\Phi_{\text{max}}-\Phi_{\text{min}}$
plane in figure \ref{fig:PhaseDiagram}. Note that according to the
KS equation \eqref{eq:ksgl}, the observed patterns should only depend
on the effective curvature coefficient $\tilde{\nu}$ given by equation
\eqref{eq:EffectiveCurvatureCoefficient}. However, numerical simulations
show that the type of patterns depends not only on the difference
of $\Phi_{\text{max}}$ and $\Phi_{\text{min}}$, but also display
a slight dependence on their absolute values. This dependence is due
to nonlinear corrections in the relation for the one-dimensional velocity
$c$ over $\Phi$ and higher order effects neglected by the KS equation
\eqref{eq:ksgl}.\\
\begin{figure}[t]
\begin{center}\includegraphics{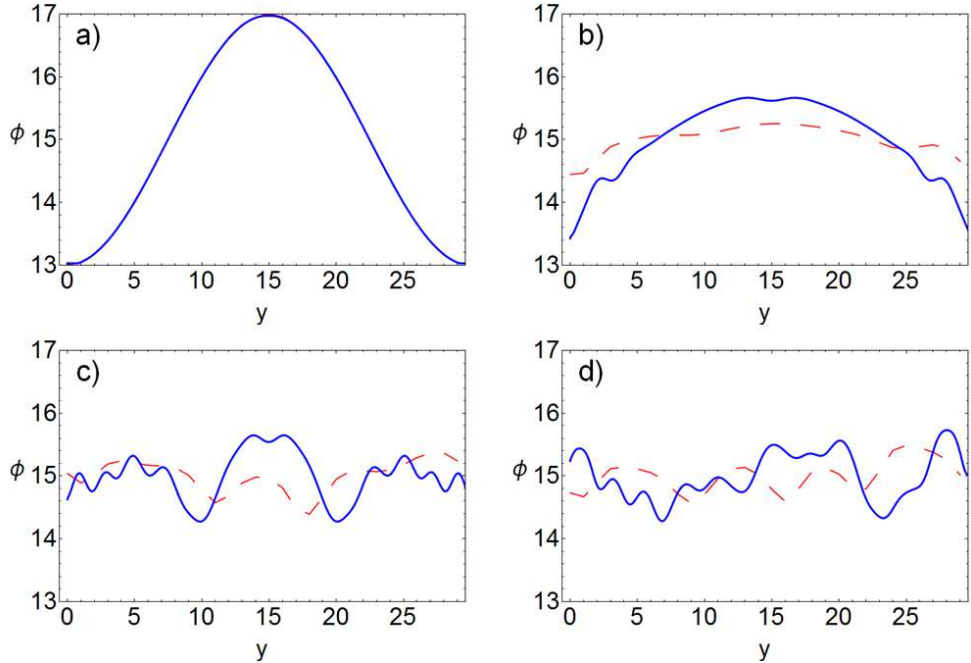}\end{center} \caption{\label{fig:TimeEvolutionComparisonKSOregonator}Time evolution of
a wave's isoconcentration line slightly beyond the threshold of transversal
instability caused by feedback with an effective curvature coefficient
$\tilde{\nu}=-0.02$. Compared are the results of two-dimensional
numerical simulations of the modified Oregonator model (dotted red
line) with the solution of the Kuramoto-Sivashinsky equation (blue
line). Due to the strong dependence on the initial conditions, the
time evolutions are comparable only for a short time span. (a) corresponds
to $t=0$, (b) $t=6$, (c) $t=12$ and (d) $t=25$. Comparison is
made in the comoving frame of reference because the centre of mass
velocities do not agree.}
\end{figure}
By adjusting the effective curvature coefficient $\tilde{\nu}$, we
are able to validate the predicted onset of transversal instabilities
equation \eqref{eq:ConditionOfTransversalInstability}, 
\begin{equation}
L=\frac{1}{\sqrt{-\tilde{\nu}}}\sqrt{\lambda}\pi\label{eq:LOversqrtnu}
\end{equation}
and its dependence on the channel width $L$. We perform numerical
simulations of the controlled Oregonator model in a channel with width
$L$ and Neumann boundary conditions in the $y$-direction. Starting
with a plane box-like initial condition equation \eqref{eq:InitialCondition}
with noisy box width $b$, we change the effective curvature coefficient
$\tilde{\nu}$ until a plane wave becomes unstable, i.e., the curvature
along the isoconcentration line is different from zero. Figure \ref{fig:StabilityThreshold}
shows that both numerical simulations and analytical prediction yield
a linear relation between channel width $L$ and $1/\sqrt{-\tilde{\nu}}$
over a large range of effective curvature coefficients $\tilde{\nu}$.
The slopes differ due to higher order corrections for the KS equation
\eqref{eq:ksgl} and nonlinear corrections for the velocity $c$ over
$\Phi$, equation \eqref{eq:c0OverPhi}, used for the feedback law.\\
Beyond the onset of transversal instabilities, the emerging patterns
can in principle be described by the KS equation \eqref{eq:ksgl}.
We compare the time evolution of the modified Oregonator model with
the solution of the KS equation for an effective curvature coefficient
of $\tilde{\nu}=-0.02$. Because the centre of mass velocity is incorrectly
predicted by the KS equation, figure \ref{fig:TimeEvolutionComparisonKSOregonator}
shows a sequence of snapshots of isoconcentration lines in a frame
of reference comoving with the centre of mass. Due to the strong dependence
on the initial data, any initial agreement between the two curves
is vanishing fast during the time evolution.

\subsection{Suppression of transversal instabilities}

\begin{figure}[h!]
\centering \includegraphics[width=1\textwidth]{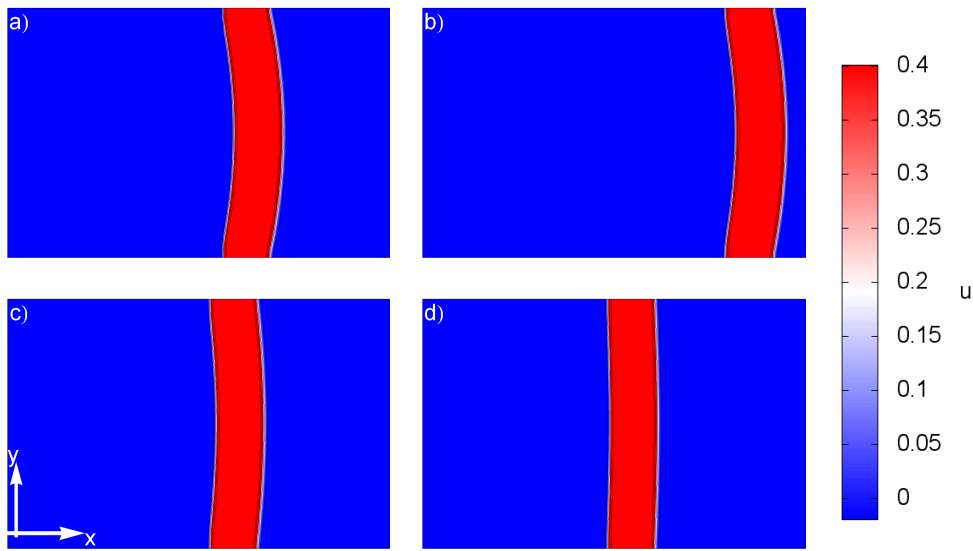} \caption{\label{fig:FHNSupressInstabilityCloseToOnset}Curvature-dependent
feedback control stabilises an unstable plane wave. For the same parameters
slightly beyond the threshold of transversal instabilities as the
corresponding uncontrolled time evolution in figure \ref{fig:FHNCloseToOnset},
the initially sinusoidally shaped wave relaxes back to a plane wave
and no fold develops. Snapshots of a movie \cite{supplement} with
(a) $t=43$, (b) $t=67$, (c) $t=159$, and (d) $t=400$. The values
of the feedback parameter are $a_{\text{min}}=0.05$ and $a_{\text{max}}=0.15$
and the control is switched on at $t=40$.}
\end{figure}

\begin{figure}[h!]
\centering \includegraphics[width=1\textwidth]{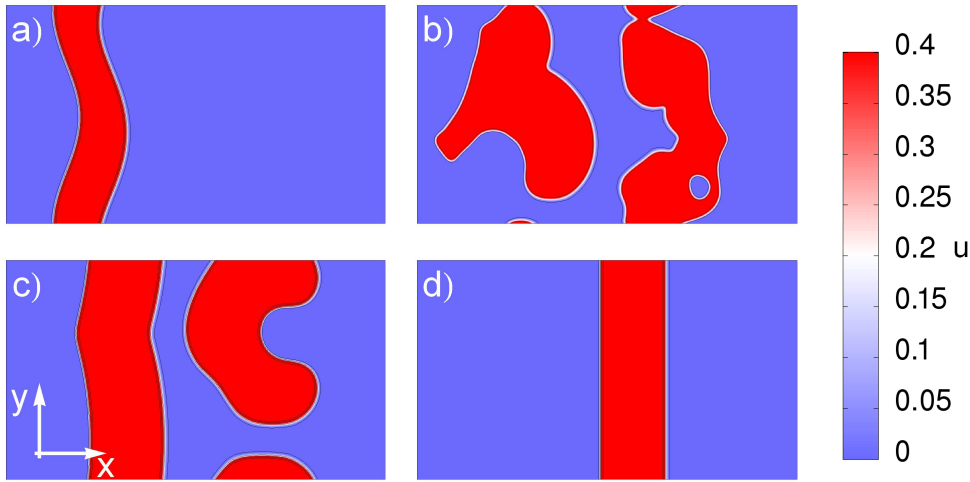} \caption{\label{fig:FHNSupressInstabilityFarFromOnset}Curvature-dependent
feedback suppresses spiral turbulence. After the feedback control
is switched on at $t=256$ (b), waves stop to break up, leaving behind
a plane wave and a pair of counter rotating spiral waves (c) and finally
a solitary plane wave (d). Same parameters deep in the regime of transversal
instabilities as the corresponding uncontrolled time evolution in
figure \ref{fig:FHNFarFromOnset}. Snapshots of a movie \cite{supplement}
with (a) $t=5$, (b) $t=301$, (c) $t=672$, and (d) $t=891$. The
values of the feedback parameter are $a_{\text{min}}=0.04$ and $a_{\text{max}}=0.08$.
}
\end{figure}

The curvature-dependent feedback law \eqref{eq:FeedbackLaw} is used
to suppress transversal instabilities occurring in the uncontrolled
piecewise linear FHN model given by equations \eqref{eq:PiecewiseLinearFHNActivator},
\eqref{eq:PiecewiseLinearFHNInhibitor}. Here, the excitation threshold
$a$ is used as the feedback parameter. First, we linearly approximate
the velocity - excitation threshold relation as
\begin{eqnarray}
c\left(a\right) & = & c_{0}+c_{1}a,
\end{eqnarray}
with $c_{0}=2.23$ and $c_{1}=-8.75$. Similar as shown in figure
\eqref{fig:c0OverPhi} for the modified Oregonator model, solitary
pulses exist only for a certain range of $a$ values. Second, the
dependence of the excitation threshold $a$ on the curvature $\kappa$
is chosen as
\begin{eqnarray}
a\left(\kappa\right) & = & \left\{ \begin{array}{@{}l@{\quad}l}
a_{\text{min}}+\beta\left(t\right)\kappa, & \kappa\geq0,\\
a_{\text{min}} & \kappa<0.
\end{array}\right.\label{control1}
\end{eqnarray}
The coefficient $\beta\left(t\right)$ is adjusted in time such that
the maximum value of $a\left(\kappa\right)$ along the isoconcentration
line does not exceed or undershoot the range of existence of solitary
pulses. Every 100 time steps, we determine the maximum curvature $\kappa_{\text{max}}\left(t\right)$
along the isoconcentration line and set $\kappa_{\text{max}}$ to
this value, 
\begin{eqnarray}
\beta\left(t\right) & = & \frac{a_{\text{max}}-a_{\text{min}}}{\kappa_{\text{max}}\left(t\right)}.
\end{eqnarray}
The background value of $a$ is set to $a_{0}=0.1$ everywhere before
the feedback control is switched on at time $t=t_{1}=215$, and outside
the region affected by the feedback control. Figure \ref{fig:FHNSupressInstabilityCloseToOnset}
displays the suppression of a transversal instability slightly beyond
the threshold. For the same parameter values as in figure \ref{fig:FHNCloseToOnset},
the initially sinusoidally shaped wave relaxes back to a plane wave
and no fold appears, see also the video in the supplemental material
\cite{supplement}.
 Patterns deep in the regime of transversal instabilities are characterised
by a continuing segmentation of waves and spreading spiral turbulence
as shown in figure \ref{fig:FHNFarFromOnset}. For the same parameter
values, patterns stop to segment after the feedback is switched on,
giving rise to a persistent plane wave and two counter rotating spiral
waves, see figure \ref{fig:FHNSupressInstabilityFarFromOnset}. The
wave front of rotating patterns has a nonbinding positive curvature.
According to the linear eikonal equation \eqref{eq:LinearEikonalEquation},
it advances slower than a plane wave if the effective curvature coefficient
$\tilde{\nu}$ is positive. Therefore, the plane wave has a tendency
to annihilate rotating waves, finally leading to a solitary plane
wave.

\section{Conclusions}

In this article, we present a feedback loop to induce, control, and
suppress transversal instabilities of reaction-diffusion waves. The
control signal is calculated from the local curvature of the isoconcentration
line of the wave. We show that the curvature dependent control can
amplify or quench small curvature perturbations in the wave shape.
Simultaneously, the feedback allows to study a large variety of artificially
produced wave patterns associated with transversal instabilities.
Often these patterns are non-stationary and sensitively depending
on small changes in the initial conditions characteristic for chaotic
dynamics.\\
Mathematically, the onset of transversal instabilities can be understood
with the help of the linear eikonal equation which relates the wave
velocity normal to an isoconcentration line to its local curvature.
The coefficient $\nu$ in front of the curvature determines the stability
of a flat wave. For positive values of $\nu$, convex wave segments
slow down while concave wave segments propagate at a higher velocity.
Under these conditions a perturbed flat traveling wave recovers its
flat shape. In the case of negative $\nu$, a small positive curvature
causes an increase of the wave velocity which in turn results in an
increase of the local curvature. Now, a flat wave is unstable with
respect to small curvature perturbations. The proposed feedback loop
is able to change the sign of the coefficient $\nu$.\\
With experiments on chemical waves in the PBZR in mind, for realistic
parameter values we show in numerical simulations with the Oregonator
model that transversal instabilities of planar waves can be induced
by the feedback. Right beyond the transversal instability of planar
waves, we find nearly flat folded waves which are stationary in a
comoving frame of reference. For weak feedback we observe small ripple-shaped
undulations traveling along the wave front. Upon increasing the feedback
strength further, V-shaped wave patterns with spatio-temporal transversal
modulations appear. These V-shaped waves travel at a velocity that
depends on the opening angle but is considerably faster than that
of the planar wave. Far away from the instability threshold, breakup
of waves causes persistent annihilation and merging of excited domains,
self-sustained rotatory motion and nucleation of rotating wave segments.
Qualitatively, the emerging wave patterns correspond to those observed
in numerical simulations with separated activator and inhibitor diffusivity
\cite{zykov1998wave}.\\
Regarding chemical wave propagation in the PBZR, we emphasise that
the feedback parameters of the control law are experimentally accessible.
For appropriate BZ recipes the dependence of the wave velocity on
the intensity of applied light should be strong enough to induce transversal
wave instabilities. The isoconcentration line of the wave can be determined
by 2d spectrophotometry with sufficient spatial resolution using the
contrast between the oxidised and reduced form of the catalyst. We
believe that the computation of the curvature by the Level Set Method
as described in Sec. \ref{sub:ComputationOfCurvature} will work reliably
for noisy experimental data, too. Because all chemical components
share similarly shaped isoconcentration lines, the measurement of
the concentration field of an arbitrary single chemical species is
sufficient for setting up the control loop. Fine-tuning the feedback
parameters allows to study the onset of transversal instabilities
in dependence of the boundary conditions as e.g. the channel width
$L$, as pointed out in chapter Sec. \ref{sub:EvolutionEquations}.\\
In the opposite case, sufficiently strong feedback changes the sign
of the effective curvature coefficient from negative to positive.
Consequently, naturally occurring transversal wave instabilities leading
to the breakup of waves are suppressed - the feedback stabilises planar
waves and spiral waves. Spreading of spiral turbulence is inhibited
due to the suppression of segmentation of waves.\\
Reaction-diffusion waves describe, at least approximately, a huge
variety of wave processes in biology. Our results are potentially
applicable to deliberately induce or inhibit transversal wave instabilities
and to control the emerging patterns under very general conditions.
The essential condition for applicability is that the propagation
velocity of the wave can be externally controlled over a sufficiently
large range such that the curvature coefficient of the eikonal equation
switches its sign.\\
Moreover, we expect that curvature dependent feedback might have interesting
applications in interfacial pattern formation in general. For example,
this feedback mechanism could be the starting point for a control
strategy aimed at the purposeful selection of patterns affected by
interfacial instabilities as, e.g., alloys growing into an undercooled
melt.\\
\textcolor{red}{}

\ack{}{We acknowledge financial support from the German Science Foundation
(DFG) within the GRK 1558 (J. L.) and within the framework of Collaborative
Research Centre 910 (S. M. and H. E.).}

\appendix

\section{\label{sec:AppendixA}Parameter values for numerical simulations}

\begin{table}[H]
\centering %
\begin{tabular}{|l||c||c|}
\hline 
parameter  & value & description\tabularnewline
\hline 
\hline 
$f$  & $1.4$  & stoichiometric parameter\tabularnewline
\hline 
\hline 
$q$  & $0.002$  & system parameter\tabularnewline
\hline 
\hline 
$1/\epsilon$  & $49$  & time scale separation\tabularnewline
\hline 
\hline 
$1/\tilde{\epsilon}$  & $4410$  & time scale separation\tabularnewline
\hline 
\hline 
$\Phi_{0}$ & $0.02$ & background illumination\tabularnewline
\hline 
\hline 
$D_{u}$  & $1.0$  & activator diffusion coefficient\tabularnewline
\hline 
\hline 
$D_{w}$  & $1.2$  & inhibitor diffusion coefficient\tabularnewline
\hline 
\hline 
$\nu$ & $1.05$ & curvature coefficient\tabularnewline
\hline 
\hline 
$\lambda$ & $0.68$ & fourth order coefficient in the KS equation\tabularnewline
\hline 
\hline 
$c_{1}$ & $-90.19$ & slope of linear fit for velocity over $\Phi$\tabularnewline
\hline 
\hline 
$c_{0}$ & $9.01$ & constant of linear fit for velocity over $\Phi$\tabularnewline
\hline 
\hline 
$\kappa_{\text{norm}}$ & $1.2$ & curvature normalisation\tabularnewline
\hline 
\hline 
$\Delta t$  & $0.0001$  & time step\tabularnewline
\hline 
\hline 
$\Delta x$,$\Delta y$  & $0.05$  & step width of spatial resolution\tabularnewline
\hline 
\end{tabular}\caption{\label{tab:3cOregonatorModelTab}Parameter values used for numerical
simulations of the modified Oregonator model.}
\end{table}
\begin{table}[H]
\centering %
\begin{tabular}{|l||c||c|}
\hline 
parameter  & value  & description\tabularnewline
\hline 
\hline 
$a$  & $0.1$ & excitation threshold\tabularnewline
\hline 
\hline 
$k_{f}$  & $2$ & system parameter\tabularnewline
\hline 
\hline 
$k_{g}$  & $2$ & system parameter\tabularnewline
\hline 
\hline 
$\sigma$  & $2.1$ & ratio of diffusion coefficients\tabularnewline
\hline 
\hline 
$\delta$  & $0.01$ & system parameter\tabularnewline
\hline 
\hline 
$\epsilon$  & $0.1575$ & time scale separation\tabularnewline
\hline 
\hline 
$c_{1}$ & $-8.75$ & slope of linear fit for velocity over $a$\tabularnewline
\hline 
\hline 
$c_{0}$ & $2.23$ & constant of linear fit for velocity over $a$\tabularnewline
\hline 
\end{tabular}\caption{\label{tab:RinzelKellertab}Parameter values used for numerical simulations
of the piecewise linear FitzHugh-Nagumo model.}
\end{table}

\section{\label{sec:AppendixB}Bonnet's formula}

We prove the formula of Bonnet, i.e., we demonstrate that evaluating
the curvature field defined by equation \eqref{eq:curvaturefield}
at an isoconcentration line $\boldsymbol{\gamma}$ yields the curvature
of $\boldsymbol{\gamma}$.\\
Let $u=u\left(x,y\right)$ be the map $\mathbb{R}^{2}\rightarrow\mathbb{R}$
and $\boldsymbol{\gamma}\left(y\right)=\left(\gamma_{x}\left(y\right),y\right)^{T}$
be the isoconcentration line $\boldsymbol{\gamma}$ parametrised by
$y$. It follows that $u\left(\gamma_{x}\left(y\right),y\right)=u_{c}=\text{const}.$
for all values of $y$. Therefore we can write (to shorten the notation,
we write $\mbox{{\ensuremath{\partial_{x}}u\ensuremath{\left(x,y\right)|_{x=\gamma_{x}\left(y\right)}}=\ensuremath{\partial_{x}}u\ensuremath{\left(\gamma_{x}\left(y\right),y\right)}}}$)
\begin{eqnarray}
\frac{d}{dy}u\left(\gamma_{x}\left(y\right),y\right) & =\partial_{x}u\left(\gamma_{x}\left(y\right),y\right)\gamma_{x}'\left(y\right)+\partial_{y}u\left(\gamma_{x}\left(y\right),y\right)= & 0,\label{eq:B1}
\end{eqnarray}
 and 
\begin{eqnarray}
\frac{d^{2}}{dy^{2}}u\left(\gamma_{x}\left(y\right),y\right) & = & \frac{d}{dy}\left(\partial_{x}u\left(\gamma_{x}\left(y\right),y\right)\gamma_{x}'\left(y\right)+\partial_{y}u\left(\gamma_{x}\left(y\right),y\right)\right)\nonumber \\
 & = & \partial_{x}u\left(\gamma_{x}\left(y\right),y\right)\gamma_{x}''\left(y\right)+\partial_{x}^{2}u\left(\gamma_{x}\left(y\right),y\right)\left(\gamma_{x}'\left(y\right)\right)^{2}\nonumber \\
 & + & 2\partial_{y}\partial_{x}u\left(\gamma_{x}\left(y\right),y\right)\gamma_{x}'\left(y\right)+\partial_{y}^{2}u\left(\gamma_{x}\left(y\right),y\right)\nonumber \\
 & = & 0,\label{eq:B2}
\end{eqnarray}
and generally $\frac{d^{n}}{dy^{n}}u\left(\gamma_{x}\left(y\right),y\right)=0$
with $n\in\mathbb{N},\, n>0$. The curvature field $\tilde{\kappa}$,
Eq. \eqref{eq:curvaturefield}, expressed in Cartesian coordinates
is
\begin{eqnarray}
\tilde{\kappa}\left(x,y\right)= & \frac{\partial_{y}^{2}u\left(x,y\right)\left(\partial_{x}u\left(x,y\right)\right)^{2}-2\partial_{x}u\left(x,y\right)\partial_{y}u\left(x,y\right)\partial_{x,y}u\left(x,y\right)}{\left(\left(\partial_{x}u\left(x,y\right)\right)^{2}+\left(\partial_{y}u\left(x,y\right)\right)^{2}\right)^{3/2}}\nonumber \\
 & +\frac{\partial_{x}^{2}u\left(x,y\right)\left(\partial_{y}u\left(x,y\right)\right)^{2}}{\left(\left(\partial_{x}u\left(x,y\right)\right)^{2}+\left(\partial_{y}u\left(x,y\right)\right)^{2}\right)^{3/2}}.
\end{eqnarray}
Evaluating $\tilde{\kappa}$ at the isoconcentration line yields
\begin{eqnarray}
\fl\tilde{\kappa}\left(\gamma_{x}\left(y\right),y\right)= & \frac{\partial_{y}^{2}u\left(\gamma_{x}\left(y\right),y\right)\left(\partial_{x}u\left(\gamma_{x}\left(y\right),y\right)\right)^{2}+\partial_{x}^{2}u\left(\gamma_{x}\left(y\right),y\right)\left(\partial_{y}u\left(\gamma_{x}\left(y\right),y\right)\right)^{2}}{\left(\left(\partial_{x}u\left(\gamma_{x}\left(y\right),y\right)\right)^{2}+\left(\partial_{y}u\left(\gamma_{x}\left(y\right),y\right)\right)^{2}\right)^{3/2}}\nonumber \\
 & -\frac{2\partial_{x}u\left(\gamma_{x}\left(y\right),y\right)\partial_{y}u\left(\gamma_{x}\left(y\right),y\right)\partial_{x,y}u\left(\gamma_{x}\left(y\right),y\right)}{\left(\left(\partial_{x}u\left(\gamma_{x}\left(y\right),y\right)\right)^{2}+\left(\partial_{y}u\left(\gamma_{x}\left(y\right),y\right)\right)^{2}\right)^{3/2}}.\label{eq:B4}
\end{eqnarray}
Using Eq. \eqref{eq:B1}, the denominator of Eq. \eqref{eq:B4} can
be simplified as
\begin{eqnarray}
\fl\left(\partial_{x}u\left(\gamma_{x}\left(y\right),y\right)\right)^{2}+\left(\partial_{y}u\left(\gamma_{x}\left(y\right),y\right)\right)^{2} & =\left(\partial_{x}u\left(\gamma_{x}\left(y\right),y\right)\right)^{2}+\left(-\partial_{x}u\left(\gamma_{x}\left(y\right),y\right)\gamma_{x}'\left(y\right)\right)^{2}\nonumber \\
 & =\left(\partial_{x}u\left(\gamma_{x}\left(y\right),y\right)\right)^{2}\left(1+\left(\gamma_{x}'\left(y\right)\right)^{2}\right).
\end{eqnarray}
Similarly, using Eq. \eqref{eq:B1} and Eq. \eqref{eq:B2}, the first
term of the numerator of Eq. \eqref{eq:B4} can be rewritten in the
form
\begin{eqnarray}
\fl\partial_{y}^{2}u\left(\gamma_{x}\left(y\right),y\right)\left(\partial_{x}u\left(\gamma_{x}\left(y\right),y\right)\right)^{2} & = & -\left(\partial_{x}u\left(\gamma_{x}\left(y\right),y\right)\right)^{2}\left(\vphantom{+\partial_{x}^{2}u\left(\gamma_{x}\left(y\right),y\right)\left(\gamma_{x}'\left(x\right)\right)^{2}+2\partial_{x,y}u\left(\gamma_{x}\left(y\right),y\right)\gamma_{x}'\left(y\right)}\partial_{x}u\left(\gamma_{x}\left(y\right),y\right)\gamma_{x}''\left(x\right)\right.\\
 &  & \left.+\partial_{x}^{2}u\left(\gamma_{x}\left(y\right),y\right)\left(\gamma_{x}'\left(x\right)\right)^{2}+2\partial_{x,y}u\left(\gamma_{x}\left(y\right),y\right)\gamma_{x}'\left(y\right)\right),\nonumber 
\end{eqnarray}
while the second term of the numerator of Eq. \eqref{eq:B4} can be
cast as
\begin{eqnarray}
\fl\partial_{x}^{2}u\left(\gamma_{x}\left(y\right),y\right)\left(\partial_{y}u\left(\gamma_{x}\left(y\right),y\right)\right)^{2} & = & \partial_{x}^{2}u\left(\gamma_{x}\left(y\right),y\right)\left(\partial_{x}u\left(\gamma_{x}\left(y\right),y\right)\right)^{2}\left(\gamma_{x}'\left(x\right)\right)^{2}.
\end{eqnarray}
The last term of the numerator of Eq. \eqref{eq:B4} becomes 
\begin{eqnarray}
-2\partial_{x}u\left(\gamma_{x}\left(y\right),y\right)\partial_{y}u\left(\gamma_{x}\left(y\right),y\right)\partial_{x,y}u\left(\gamma_{x}\left(y\right),y\right) & =\nonumber \\
2\left(\partial_{x}u\left(\gamma_{x}\left(y\right),y\right)\right)^{2}\partial_{x,y}u\left(\gamma_{x}\left(y\right),y\right)\gamma_{x}'\left(y\right).
\end{eqnarray}
All terms except the term proportional to $\gamma_{x}''\left(x\right)$
in the numerator cancel. We are left with
\begin{eqnarray}
\tilde{\kappa}\left(\gamma_{x}\left(y\right),y\right) & = & -\frac{\gamma_{x}''\left(y\right)}{\left(1+\left(\gamma_{x}'\left(y\right)\right)^{2}\right)^{3/2}},
\end{eqnarray}
which is exactly the curvature of a graph, see Eq. \eqref{eq:Curvature}.

\section*{References}{}

\bibliographystyle{unsrturl}
\bibliography{lit_master}

\end{document}